# Arbitrary-Order Scattering Exceptional Points in Configurable Non-Hermitian Zero-Index Materials

Yucheng Xu[a], Ling Yin[a], Yongxing Wang[b,*], Jie Luo[a,c,*]

[a]*School of Physical Science and Technology, Soochow University, Suzhou 215006, China*

[b]*Zhangjiagang Campus, Jiangsu University of Science and Technology, Zhangjiagang 215600, China*

[c]*Jiangsu Physical Science Research Center, Nanjing 210093, China*

*Email: 201900000107@just.edu.cn (Yongxing Wang); luojie@suda.edu.cn (Jie Luo)

**ABSTRACT:** Scattering exceptional points (EPs) are non-Hermitian degeneracies where the eigenvalues and eigenvectors of scattering matrices coalesce, enabling many intriguing phenomena in optical systems. Higher-order scattering EPs are particularly notable for their ultrasensitive response to perturbations, yet achieving flexible, arbitrary-order control remains challenging. Here, we propose a configurable non-Hermitian zero-index material (ZIM) network that enables arbitrary-order scattering EPs, as rigorously proved theoretically and validated numerically. Specifically, we show that in an $N$-port non-Hermitian ZIM network embedded with loss/gain dopants, the maximum achievable EP order is $N$, and the order can be flexibly tuned from 2 to $N$ or completely eliminated by adjusting the dopants. Furthermore, we compare conventional coherent perfect absorption with absorbing EPs of different orders. Although both achieve perfect absorption of all incident waves, a second-order EP already outperforms coherent perfect absorption, and higher-order EPs provide further power-law enhancement. These findings establish a pathway toward realizing arbitrary-order EPs in open scattering systems, holding significant promise for advanced sensing applications.

## INTRODUCTION

Exceptional points (EPs) are branch-point singularities in the parameter space of non-Hermitian systems, at which two or more eigenvalues and their associated eigenvectors simultaneously coalesce [1-7]. In optical systems, EPs can be broadly categorized into two related but fundamentally distinct classes: resonant EPs and scattering EPs. Resonant EPs correspond to degeneracies of the non-Hermitian Hamiltonian (or the eigenvalue problem of quasinormal modes), and are intrinsic to the modal structure of open systems, typically defined in the complex frequency plane [1-3]. Scattering EPs, by contrast, arise as degeneracies of the scattering matrix under real-frequency excitation, and manifest in physical observables such as reflection and transmission [1-3]. Although these two types of EPs originate from distinct theoretical frameworks, they both give rise to a variety of remarkable physical phenomena, including anomalous transmission or reflection [8-14], cloaking [15-17], lasing [18-21], coherent perfect absorption (CPA) [22-27], enhanced sensing capabilities [7, 28-33], and polarization manipulation [34-36].

Higher-order EPs [25, 31, 37-46], associated with the coalescence of multiple eigenvalues, provide a promising route to substantially enhance the sensitivity of optical systems to perturbations. An $N$th-order EP exhibits an $N$th-root dependence of eigenvalue splitting on perturbations [7, 28-33], making such degeneracies particularly attractive for sensing and detection applications. In systems operating near a resonant EP, the perturbation-induced frequency splitting serves as the primary observable, and higher-order resonant EPs have been shown to produce larger frequency splitting and thus greatly enhanced sensitivity [7, 28]. In principle, resonant EPs of arbitrary order can be realized by increasing the number of coupled resonators with appropriately engineered parameters, offering flexible sensing capabilities [42-45]. Very recently, ultrasensitive responses have also been reported in systems operating near higher-order scattering EPs [25], where perturbations can induce pronounced variations in output intensity at a fixed excitation frequency, particularly in the vicinity of absorbing EPs. However, despite these advances, achieving flexible control and systematic realization of scattering EPs of arbitrary order remains a significant challenge.

In this work, we propose a configurable non-Hermitian zero-index material (ZIM) network that enables the realization of scattering EPs of arbitrary order, as demonstrated through rigorous theoretical analysis and numerical validation. ZIMs [47-54], characterized by near-zero permittivity and/or permeability, support coherent interference among waves incident from multiple channels [25, 55-57], and their effective electromagnetic response can be readily tailored via embedded dopants [17, 58-62]. Specifically, we show that in an $N$-channel non-Hermitian ZIM network incorporating with loss/gain dopants, the maximum attainable order of a scattering EP is $N$, and the EP order can be flexibly tuned from 2 to $N$, or completely eliminated, by adjusting the dopants. Furthermore, we compare conventional CPA with absorbing EPs of different orders. While both enable complete absorption of incident waves, a 2nd-order EP already outperforms CPA, and higher-order EPs provide further power-law enhancement. These results establish a pathway toward realizing and controlling arbitrary-order scattering EPs in open systems, paving the way for advanced sensing functionalities.

## RESULTS

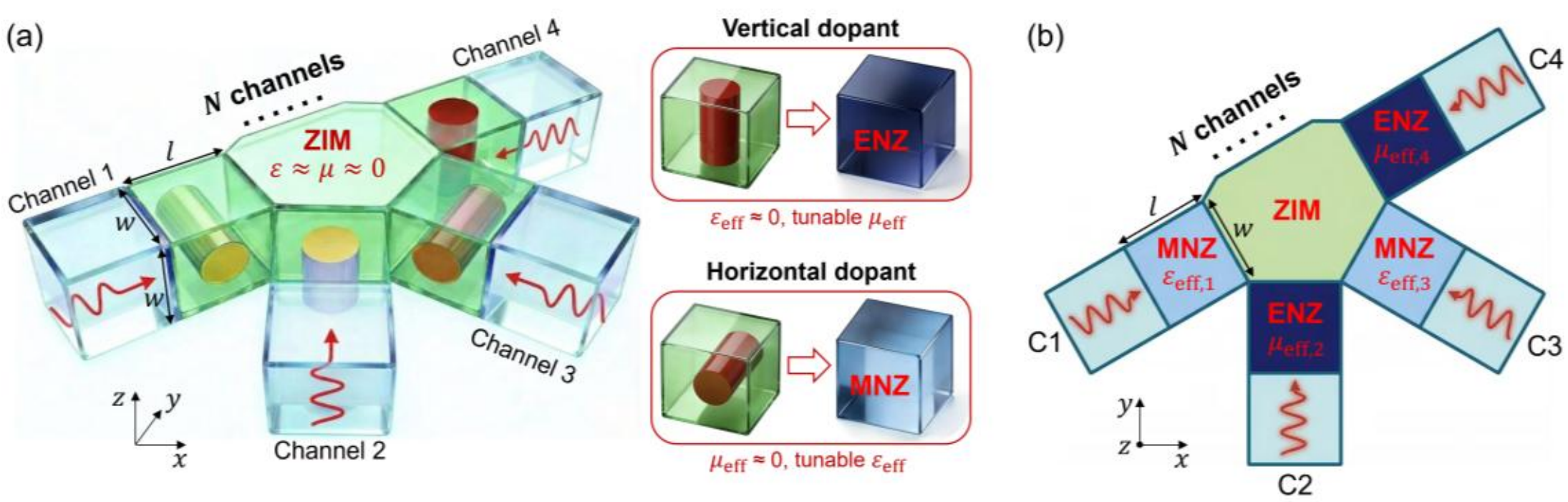

**Figure 1.** (a) Schematic of a configurable non-Hermitian ZIM network. A central ZIM is connected to $N$ channels, each comprising a hollow straight section for wave input and output, and a ZIM-filled region interfacing with the central ZIM. The ZIM components within the channels are embedded with cylindrical loss/gain dopants. These doped ZIMs can be homogenized as effective ENZ media (with $\varepsilon_{\mathrm{eff}} \approx 0$ and tunable $\mu_{\mathrm{eff}}$) or MNZ media (with $\mu_{\mathrm{eff}} \approx 0$ and tunable $\varepsilon_{\mathrm{eff}}$), depending on the orientation of dopants (insets). (b) Corresponding two-dimensional effective-medium model. The label "C" denotes a channel.

**Model and EP analysis.** Figure 1(a) illustrates the studied non-Hermitian ZIM network. It consists of a central ZIM region characterized by near-zero permittivity and permeability ($\varepsilon \approx \mu \approx 0$), which is connected to $N$ waveguide channels. This central ZIM plays a crucial role in mediating coherent interference among waves incident from different channels. Each channel has a square cross-section (side length $w$) and comprises two sections: a hollow straight section for wave input and output, and a ZIM-filled region of length $l$ that interfaces with the central ZIM. All ZIM components within the channels are identical, and they share the same material properties as the central ZIM, but are embedded with cylindrical nonmagnetic loss/gain dopants (cross-sectional radius $r_{\mathrm{d},m}$, relative permittivity $\varepsilon_{\mathrm{d},m}$ for the dopant in the $m$th channel). We note that such ZIMs with $\varepsilon \approx \mu \approx 0$ can be practically realized using photonic crystals [17, 63, 64] or metamaterials [65].

We consider transverse-magnetic polarization with magnetic field oriented along the $z$ direction and assume a time variation term of $\mathrm{e}^{-i\omega t}$, where $\omega$ is the angular frequency. The upper and lower boundaries of the network are modeled as perfect magnetic conductors (PMCs), while its side boundaries are perfect electric conductors (PECs). Although PMCs are assumed for simplicity, effective PMC boundaries can be realized experimentally using high-impedance surfaces [66-68], grooved metallic structures [69], all-dielectric metasurfaces supporting magnetic resonances [70-72], or PEC walls placed a quarter wavelength away from the ZIM [59, 73]. Under these conditions, only the fundamental transverse electromagnetic mode, characterized by the magnetic field component $H_z$, is supported in each channel. Within this framework, the doped ZIMs can be homogenized into effective media [insets of Fig. 1(a)] according to photonic doping theory [58]. Specifically, a vertically doped ZIM behaves as an effective epsilon-near-zero (ENZ) medium, with an effective relative permittivity $\varepsilon_{\mathrm{eff},m} \approx 0$ and a dopant-dependent effective relative permeability $\mu_{\mathrm{eff},m}$ expressed as [58],

$$\mu_{\mathrm{eff},m} = \frac{2\pi r_{\mathrm{d},m} J_1\left(k_0\sqrt{\varepsilon_{\mathrm{d},m}} r_{\mathrm{d},m}\right)}{k_0 l w \sqrt{\varepsilon_{\mathrm{d},m}} J_0\left(k_0\sqrt{\varepsilon_{\mathrm{d},m}} r_{\mathrm{d},m}\right)}, \tag{1}$$

where $k_0 = \omega/c$ is the free-space wavenumber, $c$ is the speed of light in vacuum, and $J_0(\dots)$ and $J_1(\dots)$ denote the zeroth- and first-order Bessel functions, respectively. In contrast, a horizontally doped ZIM can be homogenized as an effective mu-near-zero (MNZ) medium, with an effective relative permeability $\mu_{\mathrm{eff},m} \approx 0$ and a dopant-dependent effective relative permittivity $\varepsilon_{\mathrm{eff},m}$ given by [58],

$$\varepsilon_{\mathrm{eff},m} = \frac{2\pi r_{\mathrm{d},m} \sqrt{\varepsilon_{\mathrm{d},m}} J_1\left(k_0\sqrt{\varepsilon_{\mathrm{d},m}} r_{\mathrm{d},m}\right)}{k_0 l w J_0\left(k_0\sqrt{\varepsilon_{\mathrm{d},m}} r_{\mathrm{d},m}\right)}. \tag{2}$$

Detailed derivations of $\mu_{\mathrm{eff},m}$ and $\varepsilon_{\mathrm{eff},m}$ are provided in Supporting Information Section 1. The distinct homogenization behaviors for the two doping configurations can be understood from the underlying field distributions. For vertically oriented dopants (aligned with the magnetic field), the uniformity of the magnetic field within the ZIM host is preserved, while the electric field is perturbed [74], resulting in an effective ENZ medium with spatially uniform magnetic field. Conversely, for horizontally oriented dopants (aligned with the electric field), the electric field distribution remains nearly uniform, while the magnetic field is modified, leading to an effective MNZ medium with a uniform electric field profile.

Using this effective-medium description, the system can be reduced to a two-dimensional model on the $xy$ plane [Fig. 1(b)]. The wave behavior within this two-dimensional $N$-channel ZIM network can be described by an $N \times N$ scattering matrix $\mathbf{S}$. To derive an explicit form of $\mathbf{S}$, we consider a general configuration in which $M$ channels (indexed from 1 to $M$) are filled with effective MNZ media, and the remaining $N - M$ channels (indexed from $M + 1$ to $N$) are filled with effective ENZ media, realized via orientation-controlled photonic doping. The complex amplitudes of the input and output magnetic fields are denoted by $\mathbf{a} = (a_1, a_2, \dots, a_N)$ and $\mathbf{b} = (b_1, b_2, \dots, b_N)$, respectively, and are related through $\mathbf{b} = \mathbf{Sa}$. By applying Faraday's law, Ampère's circuital law, and boundary conditions, the scattering matrix can be derived in a closed form as a rank-1 perturbation of a diagonal matrix (see Supporting Information Section 2):

$$\mathbf{S} = \mathbf{D} - \xi \mathbf{v}\mathbf{v}^T, \tag{3}$$

where $\mathbf{D} = \mathrm{diag}(1 + 2\alpha_1, 1 + 2\alpha_2, \dots, 1 + 2\alpha_N)$, $\mathbf{v} = (\alpha_1, \alpha_2, \dots, \alpha_N)^T$, $\xi = 2/(\sum_{m=1}^{M} \alpha_m - (N - M) + i\beta)$, $\alpha_m = 1/(ik_0\varepsilon_{\mathrm{eff},m} l - 1)$ when $(0 < m \leq M < N)$, $\alpha_m = -1$ when $(M < m \leq N)$, and $\beta = k_0 \sum_{m=M+1}^{N} \mu_{\mathrm{eff},m}\, l$. Here, $\varepsilon_{\mathrm{eff},m}$ (or $\mu_{\mathrm{eff},m}$) is the relative permittivity (or permeability) of the effective MNZ (or ENZ) medium in the $m$th channel. The parameters $\alpha_m$ and $\beta$, governed by $\varepsilon_{\mathrm{eff}}$ and $\mu_{\mathrm{eff}}$, respectively, characterize the effective response of the MNZ and ENZ media within the channels.

The conditions for the emergence of EPs can be derived by analyzing the characteristic polynomial of the scattering matrix $\mathbf{S}$, which is given by

$$P(s) = \xi \left[\prod_{m=1}^{M} (s - 1 - 2\alpha_m)\right] (s + 1)^{(N-M-1)} Q(s), \tag{4}$$

with

$$Q(s) = -\frac{(s-1)(N-M)}{2} + \frac{i\beta(s+1)}{2} + \sum_{m=1}^{M}(s+1)\left(\frac{\alpha_m^2}{s-1-2\alpha_m} + \frac{\alpha_m}{2}\right), \quad (5)$$

where $s$ denotes an eigenvalue of $\mathbf{S}$. A $k$th-order EP corresponds to a $k$-fold root of $P(s) = 0$ accompanied by the coalescence of the associated eigenvectors. Although $s = -1$ is an $(N - M - 1)$-fold root of $P(s) = 0$, its eigenspace has dimension $N - M - 1$; it therefore represents a diabolic point (DP) [75, 76] rather than an EP. The remaining $M + 1$ eigenvalues are determined by the equation $Q(s) = 0$. Since the rational function $Q(s)$ contains $M + 1$ independent complex parameters $\{\beta, \alpha_1, \alpha_2, \ldots, \alpha_M\}$ that can be tuned via the dopants, higher-order degeneracies can be engineered at a prescribed target eigenvalue $s_{\mathrm{EP}}$. Specifically, the conditions for $k$ $(k \leq M + 1)$ eigenvalues to coalesce at $s_{\mathrm{EP}}$ are

$$Q(s_{\mathrm{EP}}) = 0 \text{ and } \left.\frac{\partial^n Q(s)}{\partial s^n}\right|_{s=s_{\mathrm{EP}}} = 0 \ (n = 1,2, \ldots, k-1), \quad (6)$$

which constitute $k$ constraints in the $(M + 1)$-dimensional complex parameter space. For $k \leq M + 1$, these constraints can generally be satisfied for any $s_{\mathrm{EP}}$, yielding a set of solutions $\{\beta, \alpha_1, \alpha_2, \ldots, \alpha_M\}$ that correspond to specific ZIM network configurations (see Supporting Information Section 3). Under these conditions, $s_{\mathrm{EP}}$ becomes a root of multiplicity $k$, leading to the degeneracy of $k$ eigenvalues at $s_{\mathrm{EP}}$. Moreover, for these $k$ degenerate eigenvalues, their eigenvectors become linearly dependent and collapse onto a single vector as

$$\mathbf{u} = (\mathbf{D} - s_{\mathrm{EP}}\mathbf{I})^{-1}\,\mathbf{v}, \quad (7)$$

where $\mathbf{I}$ is the identity matrix. The simultaneous coalescence of eigenvalues and their eigenvectors confirms the emergence of a $k$th-order EP in an $N$-channel ZIM network under appropriate parameter tuning.

**Table 1.** Maximum achievable EP order for different configurations in an $N$-channel non-Hermitian ZIM network.

| Number of MNZ channels | Number of ENZ channels | Maximum EP order |
|---|---|---|
| **0** | $N$ | None (DP of order $N-1$) |
| $\boldsymbol{M\ (0 < M < N-1)}$ | $N-M$ | $M+1$ |
| $\boldsymbol{N-1}$ | 1 | $N$ |
| $\boldsymbol{N}$ | 0 | $N-1$ |

Equations (6) and (7) not only specify the conditions for realizing a $k$th-order EP, but also establish a general design principle. In an $N$-channel ZIM network with $M$ $(0 < M < N)$ MNZ channels and $N - M$ ENZ channels, EPs of order up to $M + 1$ can be achieved. In particular, a configuration with $M = N - 1$ supports an $N$th-order scattering EP. Table 1 summarizes the maximum achievable EP order for different configurations in an $N$-channel ZIM network. Detailed proofs are provided in Supporting Information Section 3.

We note that two special cases, not captured by Eqs. (6) and (7), require separate consideration. First, when $M = 0$ (all channels filled with effective ENZ media), no EP arises. In this case, the magnetic field is uniform across the central ZIM and all doped ZIMs within channels, imposing the constraint $1 + S_{nn} = S_{mn}$ $(m \neq n)$ on scattering matrix $\mathbf{S}$. Under this condition, the eigenvalues of $\mathbf{S}$ are $s_1 = N - 1 + \sum_{n=1}^{N} S_{nn}$ and $s_2 = s_3 = \cdots = s_N = -1$. Although $N - 1$ eigenvalues are degenerate, their eigenvectors remain linearly independent, yielding an $(N - 1)$-fold DP at $s = -1$.

This behavior can be understood intuitively. In the all-ENZ network, only the in-phase excitation, corresponding to the eigenvector $\mathbf{u} = (1,1, \dots ,1)^T$ of $\mathbf{S}$, can couple energy into the network and interact with the gain/lossy dopants [59, 60]. In contrast, all out-of-phase excitations, corresponding to the remaining $N - 1$ out-of-phase eigenstates of $\mathbf{S}$, produce a vanishing magnetic field inside the ZIM region and therefore cannot couple with the dopants, behaving analogously to dark states. Incident waves associated with these out-of-phase excitations are completely reflected and share the same eigenvalue $s = -1$. Consequently, only the in-phase eigenstate is affected by the non-Hermitian tuning, whereas the remaining $(N - 1)$-dimensional subspace remains frozen, giving rise to an $(N - 1)$-fold DP rather than an EP. Physically, these out-of-phase eigenstates cannot "see" the non-Hermitian tuning and therefore retain their Hermitian degeneracy. From this perspective, the absence of EPs originates from the field-uniformity constraint imposed by the ENZ medium, which locks the relative scattering responses among different ports. Breaking this constraint is therefore essential for enabling flexible control over individual port responses and achieving eigenvector coalescence. Unlike our previous approach [25], which introduced air gaps between the central ZIM and the channels to break the field uniformity, the present strategy achieves this by reorienting the dopants to switch ENZ media to MNZ media, thereby enabling both the emergence of EPs and flexible control over their order.

Second, when $M = N$ (all channels filled with effective MNZ media), the situation is reversed. In this case, the in-phase excitation, corresponding to the eigenvector $\mathbf{u} = (1,1,\ldots,1)^T$, becomes a dark state. The forward and backward waves satisfy $a_m = b_m$, resulting in a vanishing electric field throughout the MNZ channels. Consequently, this eigenstate is completely decoupled from the gain/loss distribution and always corresponds to the fixed eigenvalue $s = 1$ (total reflection). The remaining $N - 1$ out-of-phase modes, however, support non-zero fields inside the network and therefore remain accessible to non-Hermitian tuning. Their eigenvalues and eigenvectors can be continuously modified and ultimately driven to coalescence, yielding an EP of maximum order $N - 1$.

Thus, the all-ENZ and all-MNZ cases exhibit a striking modal duality: in the all-ENZ case, only the in-phase mode remains tunable, while the remaining $(N - 1)$-dimensional subspace is frozen; in the all-MNZ case, the in-phase mode is frozen while the remaining $N - 1$ modes remain dopant-tunable and can coalesce into a higher-order EP. Representative examples are provided in Supporting Information Section 4.

**Three-channel model.** The above theoretical analysis shows that the order of EPs in the non-Hermitian ZIM network is governed by the filling materials within the channels, or equivalently, by the orientation of the embedded dopants. This property provides a direct and versatile route for controlling the EP order. For illustration, we consider a three-channel effective-medium model as a representative example, as shown in Fig. 2. Three configurations are examined: (i) all three channels filled with ENZ media [Fig. 2(a)], (ii) two ENZ channels and one MNZ channel [Fig. 2(d)], and (iii) one ENZ channel and two MNZ channels [Fig. 2(g)]. According to the general analysis (Table 1), no EP exists in the first configuration, whereas the maximum achievable EP orders in the second and third configurations are 2 and 3, respectively.

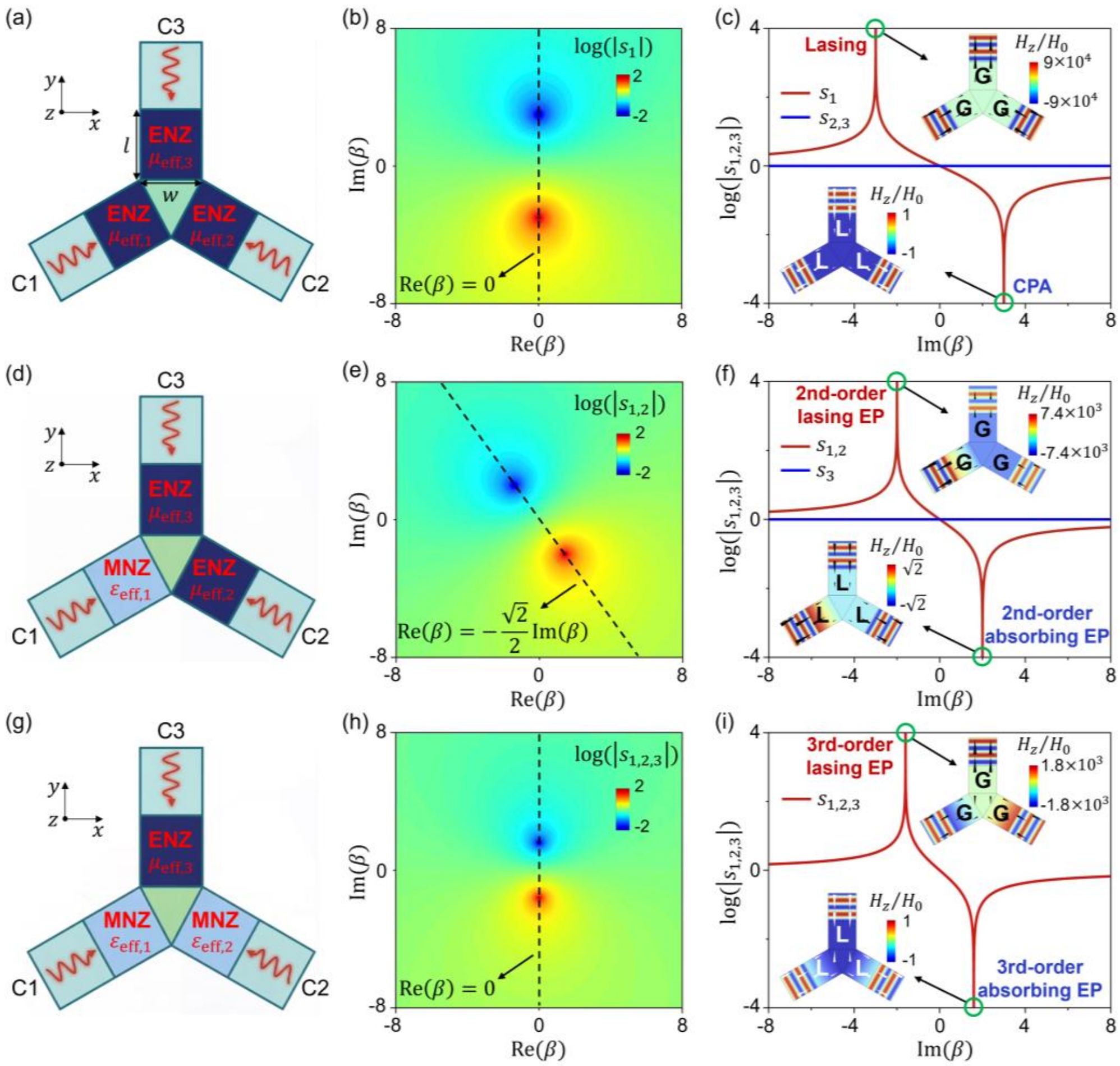

**Figure 2.** (a) Three-channel effective-medium model with all channels filled with effective ENZ media. (b) Non-degenerate scattering matrix eigenvalue $\log(|s_1|)$ in the $\{\mathrm{Re}(\beta), \mathrm{Im}(\beta)\}$ parameter space. The dashed line that passes through the pole and zero corresponds to $\mathrm{Re}(\beta) = 0$. (c) Three eigenvalues $\log(|s_{1,2,3}|)$ as a function of $\mathrm{Im}(\beta)$ along the dashed line in (b). (d) Three-channel model with two ENZ channels and one MNZ channel. (e) Degenerate eigenvalues $\log(|s_{1,2}|)$ in the $\{\mathrm{Re}(\beta), \mathrm{Im}(\beta)\}$ parameter space. The dashed line indicates $\mathrm{Re}(\beta) = -\mathrm{Im}(\beta)/\sqrt{2}$. (f) Three eigenvalues $\log(|s_{1,2,3}|)$ as a function of $\mathrm{Im}(\beta)$ along the line in (e). (g) Three-channel model with one ENZ channel and two MNZ channels. (h) Degenerate eigenvalues $\log(|s_{1,2,3}|)$ in the $\{\mathrm{Re}(\beta), \mathrm{Im}(\beta)\}$ parameter space. The dashed line corresponds to $\mathrm{Re}(\beta) = 0$. (i) Three eigenvalues $\log(|s_{1,2,3}|)$ as a function of $\mathrm{Im}(\beta)$ along the line in (h). Insets in (c), (f), and (i) show the simulated distributions of normalized magnetic field $H_z/H_0$ (color map) and time-averaged Poynting vectors (arrows) at their corresponding lasing point (scattering pole) and the CPA point or absorbing EP (scattering zero). The labels "L" and "G" denote the presence of loss or gain in the ENZ/MNZ media, respectively.

To verify these predictions, we derive the corresponding scattering matrices for the three configurations as

$$\mathbf{S} = \frac{-2}{i\beta-3}\begin{pmatrix} 1 & 1 & 1 \\ 1 & 1 & 1 \\ 1 & 1 & 1 \end{pmatrix} - \mathbf{I}, \tag{8}$$

$$\mathbf{S} = \frac{2}{i\beta+\alpha_1-2}\begin{pmatrix} (1+\alpha_1)(i\beta-2)+\alpha_1 & \alpha_1 & \alpha_1 \\ \alpha_1 & -1 & -1 \\ \alpha_1 & -1 & -1 \end{pmatrix} - \mathbf{I}, \tag{9}$$

and $$\mathbf{S} = \frac{2}{i\beta+\alpha_1+\alpha_2-1}\begin{pmatrix} (1+\alpha_1)(i\beta+\alpha_2)-1 & -\alpha_1\alpha_2 & \alpha_1 \\ -\alpha_1\alpha_2 & (1+\alpha_2)(i\beta+\alpha_1)-1 & \alpha_2 \\ \alpha_1 & \alpha_2 & -1 \end{pmatrix} - \mathbf{I}, \tag{10}$$

where $\alpha_1 = \frac{1}{ik_0\varepsilon_{\text{eff},1}l-1}$, $\alpha_2 = \frac{1}{ik_0\varepsilon_{\text{eff},2}l-1}$, and $\beta = k_0 \sum_{m=M+1}^{3} \mu_{\text{eff},m}\, l$ with $M$ denoting the number of MNZ channels. As $\beta$ depends on $\mu_{\text{eff}}$ of the effective ENZ media, its imaginary part reflects material loss ($\text{Im}(\beta) > 0$) or gain ($\text{Im}(\beta) < 0$).

For the first configuration, the scattering matrix $\mathbf{S}$ in Eq. (8) yields one tunable eigenvalue $s_1 = \frac{3+i\beta}{3-i\beta}$ and two fixed eigenvalues $s_{2,3} = -1$. The eigenvectors associated with the degenerate eigenvalue $s_{2,3} = -1$ remain linearly independent, indicating a DP rather than an EP; hence no EP exists. The expression for $s_1$ reveals two scattering singularities: a scattering zero ($s_1 = 0$) at $\beta = 3i$ and a scattering pole ($s_1 \to \infty$) at $\beta = -3i$. For clarity, we plot $\log|s_1|$ in the $\{\text{Re}(\beta), \text{Im}(\beta)\}$ parameter space in Fig. 2(b), with structural parameters $l = w = 2\lambda_0$ ($\lambda_0$ is the free-space wavelength). As expected, a scattering zero and a scattering pole appear along the line $\text{Re}(\beta) = 0$ (dashed line), as further shown in Fig. 2(c), which plots all three eigenvalues along this line. Specifically, $s_1 = 0$ at $\text{Im}(\beta) = 3$ and $s_1 \to \infty$ at $\text{Im}(\beta) = -3$, while the other two eigenvalues remain constant. The scattering zero corresponds to CPA, whereas the scattering pole is associated with lasing [77, 78].

To further confirm the wave behavior at the two singularities, full-wave simulations are performed using the finite-element software COMSOL Multiphysics. The insets in Fig. 2(c) show the simulated distributions of normalized magnetic field $H_z/H_0$ (color map) and time-averaged Poynting vectors (arrows) at the CPA and lasing points. For the CPA configuration, the three channels are filled with MNZ media characterized by $\mu_{\text{eff},1} = \mu_{\text{eff},2} = \mu_{\text{eff},3} = i/(4\pi)$; for the lasing configuration, the corresponding parameters are $\mu_{\text{eff},1} = \mu_{\text{eff},2} = \mu_{\text{eff},3} = -i/(4\pi)$. In both cases, the incident excitation is identical, with complex magnetic-field amplitudes given by the eigenvector $\mathbf{u} = H_0(1,1,1)^T$. At the CPA point, the Poynting vectors are directed inward, indicating complete absorption of the incident waves. By contrast, at the lasing point, the Poynting vectors point outward, signifying pronounced energy emission consistent with lasing behavior.

For the second configuration, the scattering matrix $\mathbf{S}$ in Eq. (9) exhibits two variable eigenvalues $s_1$ and $s_2$, and one fixed eigenvalue $s_3 = -1$. Interestingly, based on Eqs. (6) and (7), two distinct parameter sets can realize a 2nd-order EP with $s_1 = s_2 \equiv s_{1,2}$ under the condition

$$(1 + \alpha_1 + \frac{2-\alpha_1^2}{i\beta+\alpha_1-2})^2 + \frac{8\alpha_1^2}{(i\beta+\alpha_1-2)^2} = 0. \tag{11}$$

Here, we focus on one representative solution

$$s_{1,2} = \frac{-18+(-4\sqrt{2}+2i)\beta+(-1+2\sqrt{2}i)\beta^2+i\beta^3}{-18+(2\sqrt{2}+14i)\beta+5\beta^2-i\beta^3}, \tag{12}$$

which yields an absorbing EP ($s_{1,2} = 0$) at $\beta = -\sqrt{2} + 2i$ and a lasing EP ($s_{1,2} \to \infty$) at $\beta = \sqrt{2} - 2i$. These singularities are evident in the map of $\log|s_{1,2}|$ in the $\{\mathrm{Re}(\beta), \mathrm{Im}(\beta)\}$ parameter space [Fig. 2(e), with $l = w = 2\lambda_0$]. When plotting this map, $\alpha_1$ is varied continuously according to Eq. (11) to maintain the EP condition. The two singular EPs lie along the line $\mathrm{Re}(\beta) = -\mathrm{Im}(\beta)/\sqrt{2}$ (dashed line), as illustrated in Fig. 2(f), which shows all three eigenvalues along this trajectory. A fixed eigenvalue $s_3$ and two coalesced eigenvalues $s_{1,2}$ are observed, varying from zero to infinity, corresponding to 2nd-order absorbing and lasing EPs, respectively. The insets in Fig. 2(f) confirm the wave behaviors at the two singular EPs under the excitation of their corresponding eigenvector $\mathbf{u} = H_0(\sqrt{2}i, 1,1)^T$. Here, the electromagnetic parameters are $\mu_{\mathrm{eff},2} = \mu_{\mathrm{eff},3} = \varepsilon_{\mathrm{eff},1} = (-\sqrt{2} + 2i)/(8\pi)$ for the absorbing EP, with opposite values for the lasing EP. At the absorbing EP, all the incident waves are completely absorbed, whereas at the lasing EP, pronounced energy emission is observed.

For the third configuration, all three eigenvalues of $\mathbf{S}$ in Eq. (10) are tunable, allowing their coalescence into a 3rd-order EP according to Eqs. (6) and (7). In this case, the relations among $\alpha_1$, $\alpha_2$, and $\beta$, as well as the expression for the degenerate eigenvalues $s_{1,2,3}$ are considerably complicated and cannot be expressed in a compact analytical form. We therefore numerically plot the degenerate eigenvalues $\log(|s_{1,2,3}|)$ in the $\{\mathrm{Re}(\beta), \mathrm{Im}(\beta)\}$ parameter space with $l = w = 2\lambda_0$ [Fig. 2(h)]. Simultaneously, both $\alpha_1$ and $\alpha_2$ are numerically adjusted according to the $(M+1)$ nonlinear equations in Eq. (6) to satisfy the EP condition. A scattering zero ($s_{1,2,3}=0$) and a scattering pole ($s_{1,2,3}\to\infty$) are observed, corresponding to 3rd-order absorbing and lasing EPs, respectively. These singularities occur along the line $\mathrm{Re}(\beta)=0$ (dashed line), as shown in Fig. 2(i), which plots $\log(|s_{1,2,3}|)$ along this line. The 3rd-order absorbing EP appears at $\beta=1.596i$ and the 3rd-order lasing EP occurs at $\beta=-1.596i$. Simulations under excitation of the eigenvector $\mathbf{u}=H_0(0.823\mathrm{e}^{1.2i},0.823\mathrm{e}^{-1.2i},1)^T$ further verify the wave behaviors at the two EPs, as shown by the insets of Fig. 2(i). In simulations, we set $\mu_{\mathrm{eff},3}=0.399i/\pi$ and $\varepsilon_{\mathrm{eff},1}=-\varepsilon^*_{\mathrm{eff},2}=(-0.283+0.140i)/\pi$ for the absorbing EP, with opposite values for the lasing EP.

These results demonstrate that both the emergence and the order of EPs in non-Hermitian ZIM networks can be systematically controlled by choosing different filling materials within the channels. Notably, although the CPA in Fig. 2(a) and absorbing EPs in Figs. 2(d) and 2(g) exhibit similar wave behavior, as observed in insets of Figs. 2(c), 2(f), and 2(i), their responses to perturbations differ markedly, as discussed in the following. Importantly, both the CPA and absorbing EPs can be realized using purely passive components, as evidenced by the positive imaginary part of the required parameters $\mu_{\mathrm{eff}}$ and $\varepsilon_{\mathrm{eff}}$ (labelled "L" in simulation results). This feature makes them particularly attractive for practical sensing applications.

**Photonic-doped implementation.** In the following, we show that switching between effective ENZ and MNZ media within each channel can be readily achieved by reorienting the embedded dopants, thereby providing a practical and versatile route for controlling the EP order. Although the required complex or purely imaginary values of $\varepsilon_{\mathrm{eff}}$ and $\mu_{\mathrm{eff}}$ may appear extreme, they can be realized via the photonic doping approach described by Eqs. (1) and (2) [58]. With this approach, nearly arbitrary complex values of $\varepsilon_{\mathrm{eff}}$ and $\mu_{\mathrm{eff}}$ can, in principle, be engineered using appropriately designed nonmagnetic dopants with tailored loss/gain in permittivity [17, 25, 59].

For demonstration, we construct three-dimensional photonic-doped models to realize CPA and absorbing EPs, as shown in Figs. 3(a)-3(c), corresponding to the two-dimensional effective-medium models in Figs. 2(a), 2(d), and 2(g), respectively. The geometrical parameters $l$ and $w$ are the same as in the effective-medium models, and all dopants have identical radius $r_{\mathrm{d}}=$ $0.5\lambda_0$. Using Eqs. (1) and (2) together with the target values of $\mu_{\mathrm{eff}}$ and $\varepsilon_{\mathrm{eff}}$ extracted from Fig. 2, we determine the required dopant permittivities in each channel.

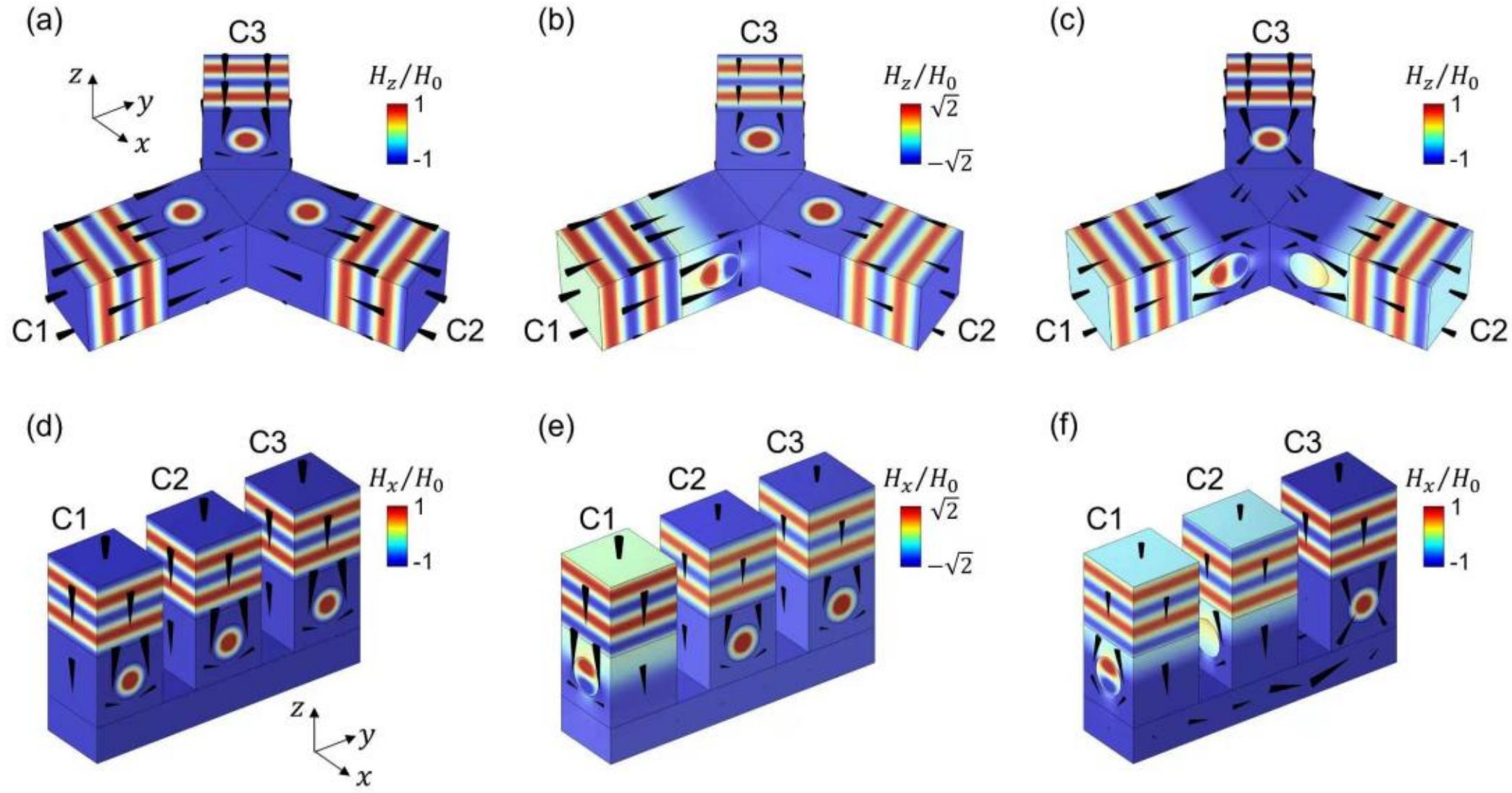

**Figure 3.** [(a)-(c)] Simulated distributions of normalized magnetic field $H_z/H_0$ (color map) and time-averaged Poynting vectors (arrows) at the CPA point or absorbing EPs for the photonic-doped models, corresponding to the effective-medium models in Figs. 2(a), 2(d), and 2(g), respectively. [(d)-(f)] Corresponding results for parallel-channel configurations, where the original triangular central ZIM in (a)-(c) is replaced with a flat geometry while all other parameters remain unchanged.

For the CPA case [Fig. 3(a)], all three dopants are identical, with $\varepsilon_{\mathrm{d},1} = \varepsilon_{\mathrm{d},2} = \varepsilon_{\mathrm{d},3} = 1.169 + 0.519i$, and are oriented vertically, such that the doped ZIM in each channel behaves as an effective ENZ medium. The simulated distributions of normalized magnetic field $H_z/H_0$ (color map) and time-averaged Poynting vectors (arrows) show complete absorption of all incident waves, consistent with the effective-medium model [inset of Fig. 3(c)]. Figures 3(b) and 3(c) present the photonic-doped realizations of the 2nd- and 3rd-order absorbing EPs, with dopant permittivities $\varepsilon_{\mathrm{d},1} = 1.115 + 0.374i$ and $\varepsilon_{\mathrm{d},2} = \varepsilon_{\mathrm{d},3} = 1.045 + 0.258i$ for Fig. 3(b), and $\varepsilon_{\mathrm{d},1} = 1.053 + 0.152i$, $\varepsilon_{\mathrm{d},2} = 0.296 + 0.076i$, and $\varepsilon_{\mathrm{d},3} = 0.885 + 0.494i$ for Fig. 3(c). Here, horizontally oriented dopants produce effective MNZ media. The simulated field and power-flow distributions agree with the effective-medium counterparts [insets of Figs. 3(f) and 3(i)], validating the photonic-doped implementations using purely lossy dopants.

Notably, the shape and area of the central ZIM do not affect the CPA or EPs, allowing substantial geometric flexibility. As shown in Figs. 3(d)-3(f), the triangular central ZIM in Figs. 3(a)-3(c) is replaced by a flat geometry, making all channels parallel, while leaving all other parameters unchanged. Simulations of these parallel-channel configurations again show perfect absorption of all incident waves. These results not only demonstrate the robustness of the photonic-doped implementations, but also highlight the flexibility of controlling EP emergence and EP order through dopant orientation in non-Hermitian ZIM networks.

To evaluate the practical feasibility of the proposed photonic-doped models, we perform a comprehensive Monte Carlo tolerance analysis by introducing random fluctuations into the key geometrical and material parameters of the models in Figs. 3(a)-3(c). We find that the higher-order EPs do not suffer from anomalous error amplification, and the photonic-doped platform maintains a reasonable degree of robustness against fabrication variations. Detailed analysis is provided in Supporting Information Section 5.

Furthermore, we evaluate the influence of material dispersion and frequency detuning by assuming linear dispersion for the model exhibiting a 3rd-order absorbing EP [Figs. 2(g)-2(i)]. We find that the characteristic cube-root eigenvalue splitting scaling remains well preserved for frequency detuning up to $0.3\%$, indicating that the platform maintains a finite operational bandwidth. Detailed results are provided in Supporting Information Section 6.

**Enhanced sensitivity.** Higher-order EPs provide a powerful route to enhancing the sensitivity of optical systems to perturbations. An $N$th-order EP exhibits an $N$th-root dependence of eigenvalue splitting on perturbation strength [7, 28-33], enabling amplified responses. Beyond the extensively studied resonant EPs [7, 28-33], higher-order scattering EPs have shown similarly enhanced sensitivity [25]. Here, we systematically compare the perturbation response of configurable non-Hermitian ZIM networks operating at the CPA point, 2nd-order, and 3rd-order absorbing EPs, corresponding to the models in Fig. 2(a), 2(d), and 2(g), respectively. These operating states can be flexibly switched by controlling the orientation of dopants within the channels. Importantly, operation at these states requires only passive components, making this platform particularly attractive for practical sensing and detection applications.

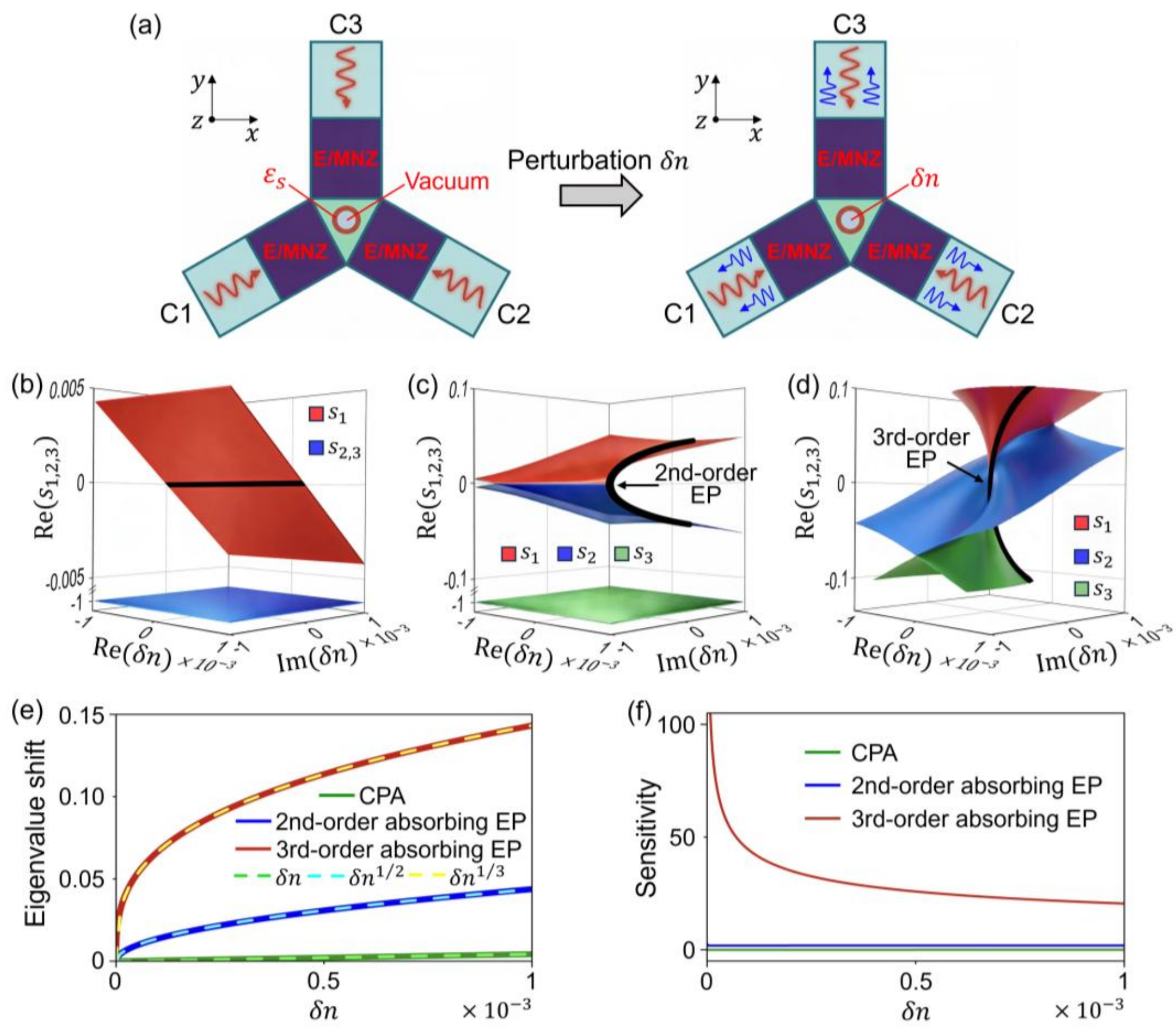

**Figure 4.** (a) Schematic of a refractive-index sensor employing a three-channel ZIM network. The central ZIM hosts a core-shell dopant (dielectric shell with hollow core) designed to keep the effective permittivity and permeability of the doped central ZIM near zero. The hollow core allows analyte gases to introduce a refractive-index perturbation $\delta n$. Without perturbation, all incident waves are perfectly absorbed at the CPA point or absorbing EPs (left). A finite perturbation $\delta n$ induces reflection, leading to a measurable change in the output power at the operating frequency (right). [(b)-(d)] Real parts of scattering matrix eigenvalues $s_{1,2,3}$ in the $\{\mathrm{Re}(\delta n),\ \mathrm{Im}(\delta n)\}$ parameter space near the (b) CPA point, (c) 2nd-order absorbing EP, and (d) 3rd-order absorbing EP. The black lines indicate the eigenvalue variation on the plane $\mathrm{Im}(\delta n) = 0$. (e) Eigenvalue shift and (f) sensitivity versus $\delta n$ (with $\mathrm{Im}(\delta n) = 0$) in the vicinity of the CPA point, 2nd-order, and 3rd-order absorbing EPs. Dashed lines in (e) represent theoretical fits to linear, square-root, and cubic-root scaling laws.

For clarity, we reconsider the three-channel effective-medium models studied in Fig. 2, with perturbations introduced via a core-shell dopant embedded in the central ZIM. The dopant consists of a dielectric shell (relative permittivity $\varepsilon_{\mathrm{s}} = 4$) and a hollow core. The hollow core serves as the sensing region, allowing analyte gases to pass through and modify the refractive index inside it. Specifically, the refractive index inside the core is expressed as $n = 1 + \delta n$, where the baseline value $n = 1$ corresponds to vacuum, and $\delta n$ denotes the refractive-index variation induced by the analyte gas. The inner and outer radii are chosen as $r_{\mathrm{c}} = 0.2925\lambda_0$ and $r_{\mathrm{s}} = 0.3235\lambda_0$, respectively, such that the effective medium of the doped central ZIM remains near-zero in both permittivity and permeability, and the unperturbed system ($\delta n = 0$) operates exactly at the CPA or absorbing EP condition. In this unperturbed case, all incident waves are perfectly absorbed [Fig. 4(a), left panel]. When analyte gases enter the hollow core, the induced refractive-index perturbation $\delta n$ shifts the scattering matrix eigenvalues away from zero, and reflection emerges, leading to a measurable change in output power at the operating frequency [Fig. 4(a), right panel]. Since $\delta n$ is directly related to the gas concentration, this mechanism can be exploited for low-concentration gas sensing [79].

We next examine the eigenvalue shift induced by $\delta n$. Figures 4(b)-4(d) show the real part of eigenvalues near the CPA point, 2nd-order, and 3rd-order absorbing EPs, respectively, in the $\{\mathrm{Re}(\delta n),\ \mathrm{Im}(\delta n)\}$ parameter space. For the CPA case, the eigenvalue $s_1$ varies linearly across zero [Fig. 4(b)]. By contrast, the EP cases exhibit markedly enhanced variations. Specifically, the 3rd-order EP [Fig. 4(d)] produces a substantially larger eigenvalue shift than the 2nd-order EP [Fig. 4(c)], as evident from the eigenvalue variation on the plane $\mathrm{Im}(\delta n) = 0$ (black lines). For a clearer comparison, we fix $\mathrm{Im}(\delta n) = 0$ (i.e., $\delta n$ is real) and plot the eigenvalue shift versus $\delta n$ [solid lines in Fig. 4(e)]. The results agree well with the theoretical linear, square-root, and cubic-root scaling laws (dashed lines), confirming the characteristic $N$th-root dependence of the eigenvalue splitting at an $N$th-order EP.

In practice, the scattering matrix eigenvalue shift is not directly measurable. Instead, the experimentally accessible quantity is the change in output power induced by the refractive-index perturbation $\delta n$. To quantify this response, we define the total reflectance $R$ as the ratio of the total output power to the total input power, which reduces to $R = |s_n|^2$ when the input is adjusted to match the eigenvector associated with the shifted eigenvalue $s_n$. The sensing performance is then characterized by the sensitivity $\Delta R/\delta n$. Figure 4(f) shows the sensitivity as a function of $\delta n$ (with $\mathrm{Im}(\delta n) = 0$) in the vicinity of the CPA point, 2nd-order, and 3rd-order absorbing EPs, clearly revealing a substantial enhancement near the 3rd-order absorbing EP. Notably, the 2nd-order EP already outperforms the CPA case, although it does not exhibit the additional power-law enhancement associated with EPs of order $n \geq 3$. Specifically, while the eigenvalue splitting follows a square-root dependence, the reflectance $R = |s_n|^2$ restores linear scaling with $\delta n$, resulting in a comparatively weaker sensitivity enhancement. These results validate the feasibility of ultrasensitive refractive-index sensing using higher-order absorbing EPs and demonstrate that, while both CPA and absorbing EPs achieve perfect absorption, absorbing EPs can provide a substantially enhanced sensing performance.

**Discussion.** Unlike previous studies that focus on perturbation-induced frequency splitting at resonant EPs or transmission-spectrum splitting near resonant EPs in coupled systems [7, 28-33], our proposed ZIM network operates at a fixed frequency and probes the perturbation-induced change in output power, which is proportional to the square of the scattering matrix eigenvalue shift. We show that, in open scattering systems described by a scattering matrix, sensitivity enhancement can likewise be achieved. More specifically, for an $N$th-order absorbing EP, the reflectance at the operating frequency scales as $R = |s|^2 \propto \delta n^{2/N}$, whereas conventional CPA yields a quadratic response $R = |s|^2 \propto \delta n^2$ (i.e., $N = 1$). Thus, even a 2nd-order EP already outperforms conventional CPA, while higher-order EPs provide further power-law enhancement. We note that the reflectance scaling ($R \propto \delta n^{2/N}$) at scattering EPs is inherently less steep than the frequency splitting at resonant EPs ($\Delta\omega \propto \delta n^{1/N}$), which we acknowledge as an inherent trade-off of the present scheme.

Compared with sensing schemes that track the frequency shift of scattering minima, such as higher-order scattering zeros or CPA [33, 78] and parity-time-symmetric CPA-lasing systems [80, 81], our approach directly measures the perturbation-induced elevation of the scattering signal from zero at a single operating frequency. This single-frequency operation may simplify the measurement requirements and reduce experimental complexity.

Although this work focuses on singular EPs (i.e., absorbing EPs with $s_{\mathrm{EP}} = 0$), our framework can be extended to non-singular EPs ($s_{\mathrm{EP}} \neq 0$) by relaxing the zero-eigenvalue condition while preserving eigenvector coalescence. Such non-singular EPs generally impose fewer constraints on system parameters, offering greater experimental flexibility. Experimentally, they can be characterized via vector network analysis to reconstruct the full scattering matrix. For sensing, the reflectance response near a non-singular EP follows $R = |s_{\mathrm{EP}}|^2 + 2\mathrm{Re}(s_{\mathrm{EP}}^* C)\delta n^{1/N} + |C|^2 \delta n^{2/N}$, where $C$ is a perturbation-dependent coefficient. The leading perturbation scales as $\delta n^{1/N}$, preserving enhanced sensitivity. However, the finite background $|s_{\mathrm{EP}}|^2$ requires baseline subtraction. By contrast, absorbing EPs provide a zero-background reflectance, which could simplify measurement and improve contrast.

Finally, it is worth emphasizing that our approach enables flexible control of scattering EP orders through orientation-controlled photonic doping, offering a distinct route towards reconfigurable non-Hermitian wave manipulation. Unlike conventional resonant-EP platforms [7, 28-33], where the EP order is determined by modal coupling configurations, our scheme allows the EP order to be tuned without altering the overall system configuration. Moreover, in contrast to our previous work on higher-order scattering EPs in non-Hermitian ZIM networks [25], where the EP order is fixed by the number of channels and relies on air gaps between the central ZIM and the ZIM channels to break field uniformity, the present approach introduces dopant orientation as a new degree of freedom for controlling both the emergence and the order of scattering EPs. Furthermore, this work establishes a rigorous theoretical framework that determines the maximum achievable EP order for arbitrary configurations, which was not addressed in the previous work [25].

## CONCLUSION

We have demonstrated arbitrary-order scattering EPs in configurable non-Hermitian ZIM networks. By controlling the orientation of dopants embedded in ZIM channels, the effective media can be switched between ENZ and MNZ behavior, providing direct and flexible control over both the emergence and the order of EPs. Theoretically, we prove that an $N$-channel ZIM network can support scattering EPs up to order $N$, with the order tunable from 2 to $N$ or completely eliminated by adjusting the dopants. These predictions are validated by numerical simulations, including photonic-doped implementations. Moreover, a systematic comparison between conventional CPA and absorbing EPs of different orders reveals that, while both achieve perfect absorption, a 2nd-order EP already outperforms CPA, and EPs of order $n \geq 3$ provide further power-law enhancement in sensing performance. These findings establish a practical platform for realizing arbitrary-order scattering EPs using dopant-controlled ZIM networks, with promising applications in ultrasensitive sensing and detection.

## METHODS

We used a commercially available finite element method simulator (COMSOL Multiphysics) to simulate the propagation of electromagnetic waves.

## ASSOCIATED CONTENT

### Supporting Information

The Supporting Information is available free of charge at

Derivation of effective parameters of doped ZIMs; Derivation of scattering matrix for the ZIM network; Analysis of higher-order scattering EPs; Modal duality between all-ENZ and all-MNZ configurations; Monte Carlo evaluation of fabrication tolerances in the ZIM networks; Frequency-detuning analysis for the ZIM networks

## AUTHOR INFORMATION

### Corresponding Authors

**Jie Luo** - *School of Physical Science and Technology, Soochow University, Suzhou 215006, China; Jiangsu Physical Science Research Center, Nanjing 210093, China*; Email: luojie@suda.edu.cn

**Yongxing Wang** - *Zhangjiagang Campus, Jiangsu University of Science and Technology, Zhangjiagang 215600, China*; Email: 201900000107@just.edu.cn

### Authors

**Yucheng Xu** - *School of Physical Science and Technology, Soochow University, Suzhou 215006, China*

**Ling Yin** - *School of Physical Science and Technology, Soochow University, Suzhou 215006,*

*China*

**Funding**

National Natural Science Foundation of China (Grant No. 12374293); Natural Science Foundation of Jiangsu Province (Grant No. BK20233001), and the Undergraduate Training Program for Innovation and Entrepreneurship, Soochow University.

**Notes**

The authors declare no competing financial interest.

Supporting Information

# Arbitrary-Order Scattering Exceptional Points in Configurable Non-Hermitian Zero-Index Materials

Yucheng Xu[a], Ling Yin[a], Yongxing Wang[b,*], Jie Luo[a,c,*]

[a]*School of Physical Science and Technology, Soochow University, Suzhou 215006, China*

[b]*Zhangjiagang Campus, Jiangsu University of Science and Technology, Zhangjiagang 215600, China*

[c]*Jiangsu Physical Science Research Center, Nanjing 210093, China*

*Email: 201900000107@just.edu.cn (Yongxing Wang); luojie@suda.edu.cn (Jie Luo)

## 1. Derivation of effective parameters of doped ZIMs

In the proposed non-Hermitian zero-index material (ZIM) network (Fig. 1 in Main Text), each ZIM component within a channel consists of a cuboid ZIM host ($\varepsilon \approx \mu \approx 0$) embedded with a cylindrical nonmagnetic loss/gain dopant. The dopant in the $m$th channel has cross-sectional radius $r_{\mathrm{d},m}$ and relative permittivity $\varepsilon_{\mathrm{d},m}$. All ZIM hosts share identical geometric parameters: length $l$ and square cross-section side length $w$. We consider transverse-magnetic polarization with the magnetic field oriented along the $z$-axis.

When the dopant in the $m$th channel is aligned along the $z$ direction, the magnetic field in the ZIM host remains spatially uniform due to the near-zero permittivity. In this case, the doped ZIM behaves as an effective epsilon-near-zero (ENZ) medium. The uniform magnetic field inside the ZIM host can be written as

$$\mathbf{H}_{\mathrm{ZIM}} = H_0 \mathbf{e}_z, \tag{S1}$$

Inside the cylindrical dopant, the magnetic field satisfies the Helmholtz equation and takes the form

$$\mathbf{H}_{\mathrm{d},m}(r) = H_{\mathrm{d},m} J_0(\sqrt{\varepsilon_{\mathrm{d},m}} k_0 r)\mathbf{e}_z, \tag{S2}$$

where $r$ is the radial distance from the dopant axis, $k_0 = \omega/c$ is the free-space wavenumber, and $c$ is the speed of light in vacuum. Continuity of tangential magnetic field at the interface $r = r_{\mathrm{d},m}$ yields $H_{\mathrm{d},m} = H_0/J_0(\sqrt{\varepsilon_{\mathrm{d},m}} k_0 r_{\mathrm{d},m})$. The presence of the dopant perturbs the magnetic flux in the ZIM host. This perturbation can be captured by introducing an effective relative permeability $\mu_{\text{eff},m}$. Applying Faraday's law over the $xy$ cross-section, we obtain

$$\oint \mathbf{E}\cdot \mathrm{d}\mathbf{l} = -i\omega\mu_0\mu_{\text{eff},m}H_0A_{xy} = -i\omega\mu_0\int_{A_{xy}} H_{\text{d},m}J_0(\sqrt{\varepsilon_{\text{d},m}}k_0r)\mathrm{d}A, \quad \text{(S3)}$$

where $A_{xy} = lw$ is the cross-section area. $J_0(\dots)$ and $J_1(\dots)$ denote the zeroth- and first-order Bessel functions, respectively. From Eq. (S3), we obtain

$$\mu_{\text{eff},m} = \frac{2\pi r_{\text{d},m}J_1(k_0\sqrt{\varepsilon_{\text{d},m}}r_{\text{d},m})}{k_0lw\sqrt{\varepsilon_{\text{d},m}}J_0(k_0\sqrt{\varepsilon_{\text{d},m}}r_{\text{d},m})}. \quad \text{(S4)}$$

On the other hand, when the dopant is aligned along the electric field direction ($y$-axis), the doped ZIM exhibits an effective mu-near-zero (MNZ) response. The electric field in the ZIM host remains approximately uniform,

$$\mathbf{E}_{\text{ZIM}} = E_0\mathbf{e}_y, \quad \text{(S5)}$$

while inside the dopant it takes the form

$$\mathbf{E}_{\text{d},m}(r) = E_{\text{d},m}J_0(\sqrt{\varepsilon_{\text{d},m}}k_0r)\mathbf{e}_y, \quad \text{(S6)}$$

with $E_{\text{d},m} = H_0/J_0(\sqrt{\varepsilon_{\text{d},m}}k_0r_{\text{d},m})$. The presence of the dopant modifies the electric flux within the ZIM host, which can be characterized by an effective relative permittivity $\varepsilon_{\text{eff},m}$. Applying Ampère's law over the $xz$ cross-section, we obtain

$$\oint \mathbf{H}\cdot \mathrm{d}\mathbf{l} = i\omega\varepsilon_0\varepsilon_{\text{eff},m}E_0A_{xz} = i\omega\varepsilon_0\varepsilon_{\text{d},m}\int_{A_{xz}} E_{\text{d},m}J_0(k_{\text{d},m}r)\,\mathrm{d}A, \quad \text{(S7)}$$

where $A_{xz} = lw$. Equation (S7) then leads to

$$\varepsilon_{\text{eff},m} = \frac{2\pi r_{\text{d},m}\sqrt{\varepsilon_{\text{d},m}}J_1(k_0\sqrt{\varepsilon_{\text{d},m}}r_{d,m})}{k_0lwJ_0(k_0\sqrt{\varepsilon_{\text{d},m}}r_{\text{d},m})}. \quad \text{(S8)}$$

## 2. Derivation of scattering matrix for the ZIM network

We consider an $N$-channel ZIM network in which each channel is loaded with a doped ZIM component that behaves as an effective ENZ or MNZ medium. We assume that channels 1 to $M$ exhibit ENZ responses, while channels $M+1$ to $N$ exhibit MNZ responses. Under transverse-magnetic illumination, the magnetic field in the $m$th channel can be expressed as

$$\mathbf{H}_m = a_n(\delta_{mn}e^{ik_0x_m} + S_{mn}e^{-ik_0x_m})\mathbf{e}_z, \quad \text{(S9)}$$

where $a_n$ is the input amplitude, $S_{mn}$ is the scattering parameter, and $x_m$ denotes the longitudinal coordinate along the $m$th channel. The corresponding electric field is

$$\mathbf{E}_m = Z_0 a_n\left(e^{ik_0 x_m} - S_{mn} e^{-ik_0 x_m}\right)\mathbf{e}_\tau, \tag{S10}$$

where $Z_0$ is the impedance of vacuum, and $\mathbf{e}_\tau$ is the unit tangential vector along the channel. Applying Faraday's law over the cross-section perpendicular to the channel direction, we obtain

$$w[(\delta_{mn} + S_{mn}) - H_c] = -ik_0 \varepsilon_{\text{eff},m} wl(\delta_{mn} - S_{mn}). \tag{S11}$$

where $H_c$ denotes the uniform magnetic field in the central ZIM region. Applying Faraday's law over the entire ZIM region yields

$$-\sum_{m=1}^{N} w\,(\delta_{mn} - S_{mn}) = ik_0 \sum_{m=1}^{N} \mu_{\text{eff},m} lw\, H_c. \tag{S12}$$

For MNZ channels ($m \leq M$), $\mu_{\text{eff},m} \approx 0$; for ENZ channels ($m > M$), $\varepsilon_{\text{eff},m} \approx 0$. Solving Eqs. (S9)-(S12), we obtain

$$S_{mn} = (1 + 2\alpha_m)\delta_{mn} - \xi\alpha_m\alpha_n, \tag{S13}$$

where $\alpha_m = 1/(ik_0\varepsilon_{\text{eff},m} l - 1)$, $\xi = 2/(\sum_{m=1}^{M}\alpha_m - (N-M) + i\beta)$, $\beta = k_0 l \sum_{m=M+1}^{N}\mu_{\text{eff},m}$. In matrix form, the scattering matrix can be written as

$$\mathbf{S} = \mathbf{D} - \xi\mathbf{v}\mathbf{v}^T, \tag{S14}$$

where $\mathbf{D} = \text{diag}(1 + 2\alpha_1, 1 + 2\alpha_2, \ldots, 1 + 2\alpha_N)$ and $\mathbf{v} = (\alpha_1, \alpha_2, \ldots, \alpha_N)^T$.

### 3. Analysis of higher-order scattering EPs

Equation (S14) shows that the scattering matrix $\mathbf{S}$ is a standard rank-1 perturbation of a diagonal matrix. Using the matrix determinant lemma, its characteristic polynomial can be written as

$$P(s) = \det(s\mathbf{I} - \mathbf{D})(1 + \xi\mathbf{v}^T(s\mathbf{I} - \mathbf{D})^{-1}\mathbf{v}), \tag{S15}$$

where $\mathbf{I}$ is the identify matrix, and $s$ is the eigenvalue of $\mathbf{S}$. For $m > M$, we have $\alpha_m = -1$, which leads to

$$P(s) = \xi\left[\prod_{m=1}^{M}(s - 1 - 2\alpha_m)\right](s+1)^{(N-M-1)}Q(s), \tag{S16}$$

with

$$Q(s) = -\frac{(s-1)(N-M)}{2} + \frac{i\beta(s+1)}{2} + \sum_{m=1}^{M}(s+1)\left(\frac{\alpha_m^2}{s-1-2\alpha_m} + \frac{\alpha_m}{2}\right). \tag{S17}$$

It follows that $s = -1$ is an $(N-M-1)$-fold root of $P(s)$. However, the corresponding eigenspace has dimension $N - M - 1$, spanned by antisymmetric vectors $\mathbf{u}_m = \mathbf{e}_{M+m} - \mathbf{e}_{M+m+1}$ for $m = 1,2,\ldots,N-M-1$, where $\mathbf{e}_n$ denotes an $N$ dimensional vector whose $n$th element are 1 and all others 0. Therefore, the degeneracy at $s = -1$ corresponds to a diabolic point (DP) rather than an exceptional point (EP).

Consequently, EPs arise from the coalescence of remaining eigenvalues determined by $Q(s) = 0$. A $k$th-order EP at $s = s_{\mathrm{EP}}$ requires that $s_{\mathrm{EP}}$ be a $k$-fold root of $Q(s) = 0$. This imposes that the $M+1$ independent complex parameters $\{\beta, \alpha_1, \alpha_2, \dots, \alpha_M\}$ simultaneously satisfy the following $k$ constraints

$$Q(s_{EP}) = 0,\ \left.\frac{\partial^n Q(s)}{\partial s^n}\right|_{s=s_{EP}} = 0\ (n = 1,2,\dots,k-1;\ k \le M+1). \tag{S18}$$

We next prove the existence of such parameters. To simplify the analysis, we introduce the rational function

$$O(s) = \frac{Q(s)}{s+1} = \left(\frac{1}{s+1} - \frac{1}{2}\right)(N-M) + \frac{i\beta}{2} + \sum_{m=1}^{M}\left(\frac{\alpha_m^2}{s-1-2\alpha_m} + \frac{\alpha_m}{2}\right), \tag{S19}$$

which shares common zeros with $Q(s)$ for $s \neq -1$. Hence, the $k$ constraints in Eq. (S18) are equivalent to

$$O(s_{\mathrm{EP}}) = 0,\ \left.\frac{\partial^n O(s)}{\partial s^n}\right|_{s=s_{\mathrm{EP}}} = 0\ (n = 1,2,\dots,k-1;\ k \le M+1). \tag{S20}$$

Equation (S20) defines a mapping from the $(M+1)$-dimensional complex parameter space $\{\beta, \alpha_1, \alpha_2, \dots, \alpha_M\}$ to a $k$-dimensional complex space. To assess whether these $k$ constraints can be satisfied simultaneously and whether parameters $\{\beta, \alpha_1, \alpha_2, \dots, \alpha_M\}$ can be found for any target $s_{\mathrm{EP}}$, we examine the Jacobian matrix of this mapping with respect to the selected set of $k$ parameters $\{\beta, \alpha_{l_1}, \alpha_{l_2}, \dots, \alpha_{l_{k-1}}\}$. The Jacobian matrix takes the form

$$\mathbf{J} = \begin{pmatrix} \frac{i}{2} & \frac{(s_{\mathrm{EP}}-1)^2}{2\left(s_{\mathrm{EP}}-1-2\alpha_{l_1}\right)^2} & \cdots & \frac{(s_{\mathrm{EP}}-1)^2}{2\left(s_{\mathrm{EP}}-1-2\alpha_{l_{k-1}}\right)^2} \\ 0 & \frac{-2\alpha_{l_1}(s_{\mathrm{EP}}-1)}{\left(s_{\mathrm{EP}}-1-2\alpha_{l_1}\right)^3} & \cdots & \frac{-2\alpha_{l_{k-1}}(s_{\mathrm{EP}}-1)}{\left(s_{\mathrm{EP}}-1-2\alpha_{l_{k-1}}\right)^3} \\ \cdots & \cdots & \cdots & \cdots \\ 0 & \frac{2(-1)^{k-1}(k-1)!\alpha_{l_1}\left[s_{\mathrm{EP}}-1+(k-2)\alpha_{l_1}\right]}{\left(s_{\mathrm{EP}}-1-2\alpha_{l_1}\right)^{k+1}} & \cdots & \frac{2(-1)^{k-1}(k-1)!\alpha_{l_{k-1}}\left[s_{\mathrm{EP}}-1+(k-2)\alpha_{l_{k-1}}\right]}{\left(s_{\mathrm{EP}}-1-2\alpha_{l_{k-1}}\right)^{k+1}} \end{pmatrix}. \tag{S21}$$

The corresponding determinant can be written as

$$\det(\mathbf{J}) = \frac{i}{2}\, C_{11}(\mathbf{J}), \tag{S22}$$

where $C_{11}(\mathbf{J})$ is the $(1,1)$-cofactor of $\mathbf{J}$. Notably, $C_{11}(\mathbf{J})$ exhibits a generalized Vandermonde-like structure. When the selected parameters are mutually distinct, the poles of these rational terms are spatially separated in the complex plane, ensuring that the rows and columns of the $C_{11}(\mathbf{J})$ remain linearly independent. Consequently, $\det(\mathbf{J})$ is generally nonzero, implying that the mapping from the $k$-dimensional parameter space $\{\beta, \alpha_{l_1}, \alpha_{l_2}, \dots, \alpha_{l_{k-1}}\}$ to the $k$-dimensional condition space is locally surjective. This guarantees that the mapping is locally non-degenerate and allows the system to reach a solution through continuous parameter tuning. Finally, $s_{\mathrm{EP}}$ become a $k$-fold root for the characteristic polynomial $P(s) = 0$ with $s_{\mathrm{EP}} \neq -1$.

For any eigenvalue $s_m$ other than $-1$, the matrix $\mathbf{D} - s_m\mathbf{I}$ is non-singular, enabling the eigenvector to be expressed as

$$\mathbf{u}_m = (\xi \mathbf{v}^T \mathbf{u}_m)(\mathbf{D} - s_m \mathbf{I})^{-1}\, \mathbf{v}. \tag{S23}$$

Here, $\xi \mathbf{v}^T \mathbf{u}_m$ is a scalar and can be omitted. When $k$ eigenvalues coalesce to $s_{\mathrm{EP}}$, the corresponding eigenvectors coalesce to the unique eigenvector

$$\mathbf{u} = (\mathbf{D} - s_{\mathrm{EP}} \mathbf{I})^{-1}\, \mathbf{v}. \tag{S24}$$

Lastly, we consider the two special cases. When $M = 0$ (all channels filled with ENZ media), the characteristic polynomial reduces to

$$P(s) = \frac{\xi}{2}(i\beta - N)\left(s + \frac{i\beta + N}{i\beta - N}\right)(s+1)^{(N-1)}. \tag{S25}$$

In this case, no EP arises. The $(N-1)$-fold root $s = -1$ corresponds to an $(N-1)$rd-order DP for antisymmetric inputs. The simple root $s = -(i\beta + N)/(i\beta - N) = N - 1 + \sum_{n=1}^{N} S_{nn}$ corresponds to the symmetric input $\mathbf{u} = (1,1,\dots,1)$.

In the case $M = N$ (all channels filled with MNZ media), the characteristic polynomial reduces to

$$P(s) = \xi\left[\prod_{m=1}^{N}(s - 1 - 2\alpha_m)\right]Q(s), \tag{S26}$$

with

$$Q(s) = \sum_{m=1}^{N} \frac{(s-1)\,\alpha_m}{2(s - 1 - 2\alpha_m)}, \tag{S27}$$

It can be verified that $s = 1$ is a simple root of $P(s)$, since

$$\left.\frac{\partial Q}{\partial s}\right|_{s=1} = \left.\sum_{m=1}^{N} \frac{\alpha_m^2}{(s-1-2\alpha_m)^2}\right|_{s=1} = \frac{N}{4} \neq 0. \tag{S28}$$

Hence, the maximum achievable EP order reduces to $N - 1$.

## 4. Modal duality between all-ENZ and all-MNZ configurations

In a multiple-channel scattering system, the eigenvector $\mathbf{u}$ of the scattering matrix represents the relative responses among different ports. The key physical distinction between all-ENZ and all-MNZ configurations lies in their different "dark modes" (i.e., excitation channels that cannot couple energy into the network) imposed by the two zero-index limits.

In the all-ENZ network, the magnetic field is constrained to remain uniform throughout the entire network. only the in-phase excitation, corresponding to the eigenvector $\mathbf{u} = (1,1,\dots,1)^T$ of the scattering matrix $\mathbf{S}$, can couple into the ZIM network and interact with the gain/lossy dopants [1, 2]. Consequently, its eigenvalue depends on the non-Hermitian configuration, as observed in Figs. 2(a)-2(c) of Main Text. In contrast, all out-of-phase excitations, corresponding to the remaining $N-1$ out-of-phase eigenstates of $\mathbf{S}$, produce a vanishing magnetic field inside the ZIM region and therefore cannot couple with the dopants. These out-of-phase eigenstates therefore cannot interact with the gain/lossy components, and the incident waves are completely reflected, yielding the same eigenvalue $s = -1$. Consequently, only the in-phase eigenstate is affected by the non-Hermitian tuning, whereas the remaining $N-1$ out-of-phase eigenvectors remain decoupled from the gain/loss distribution and share the same eigenvalue $s = -1$. This gives rise to an $(N-1)$-fold DP. Physically, these modes cannot "see" the non-Hermitian tuning and therefore retain their Hermitian degeneracy.

By contrast, in the all-MNZ network, the situation is reversed. In this case, the in-phase excitation, corresponding to the eigenvector $\mathbf{u} = (1,1,\dots,1)^T$, cannot couple energy into network. For this excitation, the forward and backward waves satisfy $a_m = b_m$, resulting in a vanishing electric field throughout the MNZ channels. Consequently, this eigenstate is completely independent of the loss/gain distribution, and its eigenvalue remains fixed at $s = 1$ (corresponding to total reflection). The remaining $N-1$ eigenstates, however, support non-zero fields inside the network and therefore remain accessible to non-Hermitian tuning. Their eigenvalues and eigenvectors can be continuously modified and ultimately driven to coalescence, giving rise to an $(N-1)$th-order scattering EP.

From this perspective, the all-ENZ and all-MNZ cases exhibit a striking modal duality: in the all-ENZ case, only the in-phase mode remains tunable, while the remaining $(N-1)$-dimensional subspace is frozen; in the all-MNZ case, the in-phase mode is frozen while the remaining $N-1$ modes remain dopant-tunable and can coalesce into a higher-order EP.

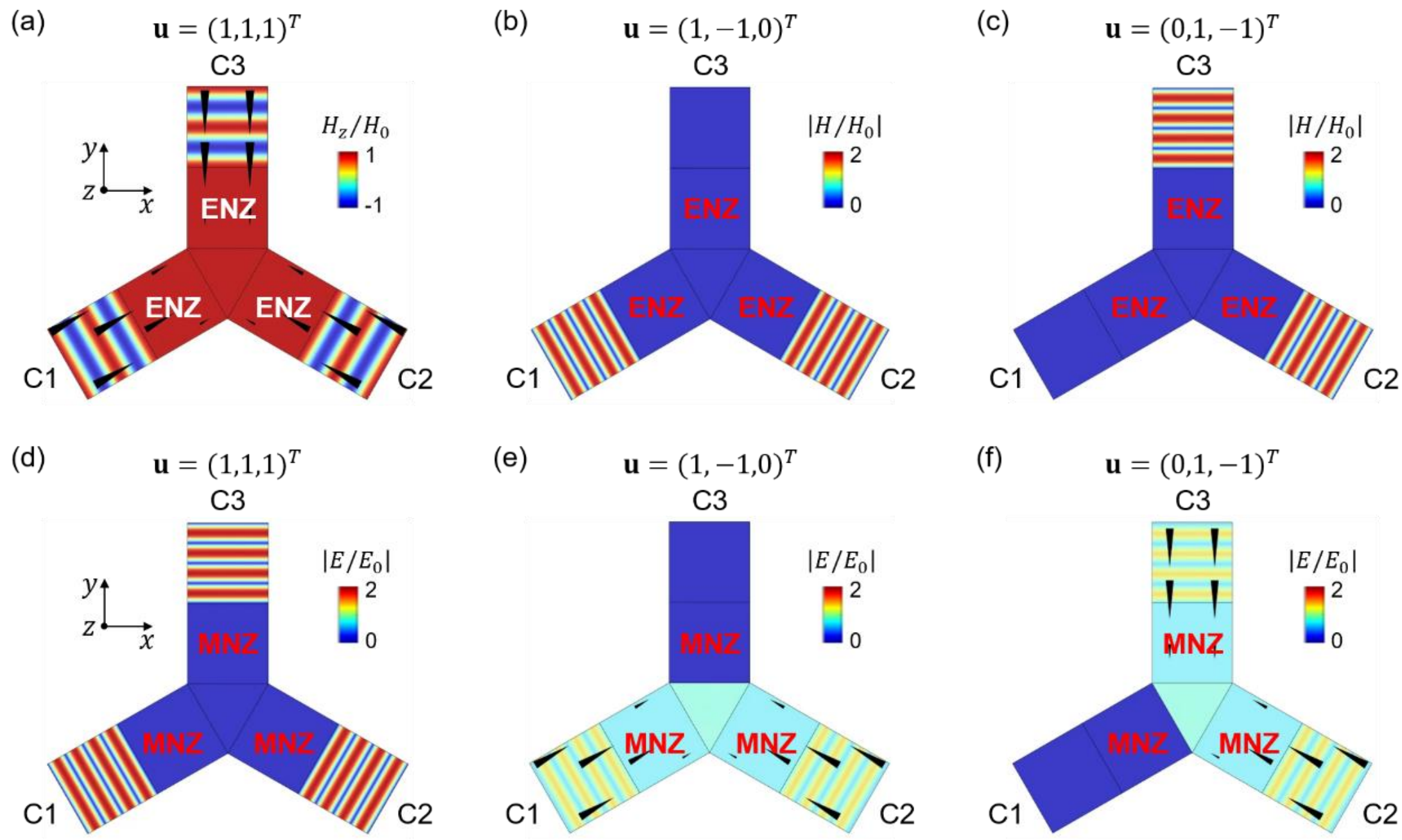

**Figure S1**. Simulated normalized field (color map) and time-averaged Poynting vectors (arrows) for the three-channel [(a)-(c)] all-ENZ and [(d)-(f)] all-MNZ networks, under excitations corresponding to the eigenvectors [(a) and (d)] $\mathbf{u_1} = (1,1,1)^T$, [(b) and (e)] $\mathbf{u_2} = (1,-1,0)^T$, and [(c) and (f)] $\mathbf{u_3} = (0,1,-1)^T$, respectively.

To further illustrate this mechanism, we take the three-channel model as a representative example, as shown in Fig. S1. For this three-channel configuration, the three eigenvectors are: $\mathbf{u_1} = (1,1,1)^T$ (in-phase eigenstate), $\mathbf{u_2} = (1,-1,0)^T$, and $\mathbf{u_3} = (0,1,-1)^T$. In the all-ENZ case, under excitation by these three eigenvectors, only the in-phase eigenstate (all incident waves with identical phase) can enter the ZIM and be affected by the loss/gain [Fig. S1(a)]. By contrast, the two out-of-phase excitations (corresponding to $\mathbf{u_2}$ and $\mathbf{u_3}$) produce a vanishing magnetic field inside the ZIM (both the central ZIM and all ENZ regions) [Figs. S1(b) and S1(c)], and thus cannot "see" the loss or gain. In the all-MNZ case, the situation is reversed: the in-phase excitation ($\mathbf{u_1}$) produces a vanishing electric field inside the ZIM (both the central ZIM and all MNZ regions) [Fig. S1(d)], whereas waves under out-of-phase excitations ($\mathbf{u_2}$ and $\mathbf{u_2}$) can enter the network and couple to the loss and gain [Figs. S1(e) and S1(f)].

## 5. Monte Carlo evaluation of fabrication tolerances in the ZIM networks

To evaluate the robustness of the proposed photonic-doped platform against fabrication imperfections, we consider the models that realize CPA and absorbing EPs in Figs. 3(a)-3(c) of the Main Text as representative examples. Specifically, we perform a Monte Carlo tolerance analysis by introducing random perturbations into three key fabrication parameters: the dopant permittivity $\varepsilon_{\mathrm{d},m}$, the dopant radius $r_{\mathrm{d},m}$, and the channel width $w$. For a given perturbation level, all corresponding parameters are independently varied using uniformly distributed random errors bounded by a prescribed maximum relative deviation. A total of 1000 independent random realizations were generated, and the resulting perturbed scattering matrix $\mathbf{S}$ is compared with the unperturbed scattering matrix $\mathbf{S}^{\mathrm{un}}$.

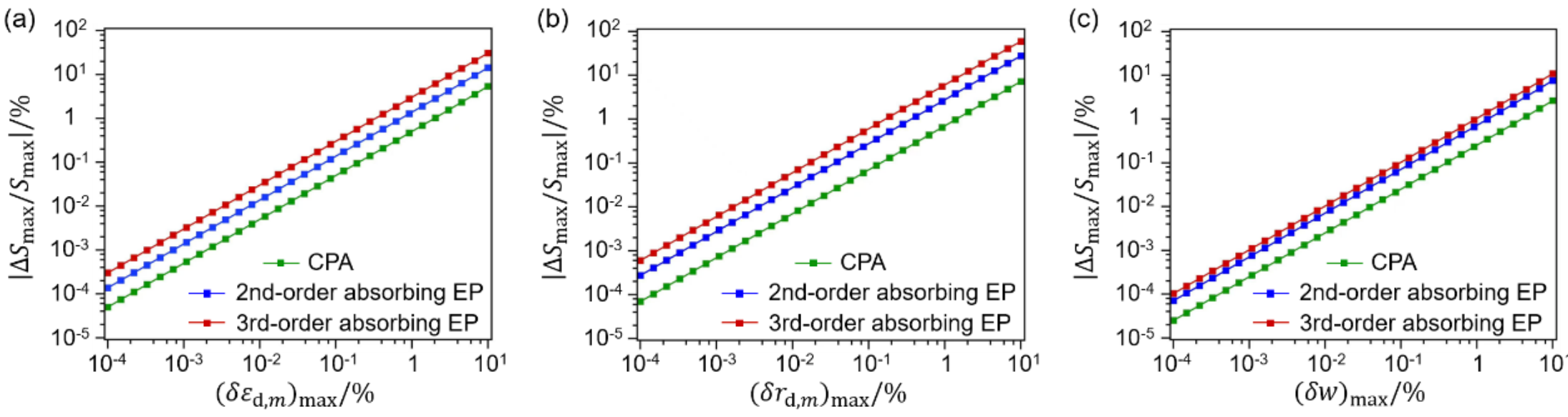

**Figure S2**. Monte Carlo tolerance analysis for the photonic-doping models in Figs. 3(a)-3(c) of the Main Text. Random perturbations are introduced into (a) the dopant permittivity $\varepsilon_{\mathrm{d},m}$, (b) the dopant radius $r_{\mathrm{d},m}$, and (c) the channel width $w$. The horizontal axis is the maximum relative fabrication uncertainty; the vertical axis shows $|\Delta S_{\max}/S_{\max}|$, the maximum absolute deviation of any scattering-matrix element between the perturbed scattering matrix $\mathbf{S}$ and the unperturbated scattering matrix $\mathbf{S}^{\mathrm{un}}$, normalized by the largest element of $\mathbf{S}^{\mathrm{un}}$. Results are presented for conventional CPA (green), 2nd-order absorbing EPs (blue) and 3rd-order absorbing EPs (red), corresponding to the models in Figs. 3(a)-3(c), respectively.

The results are summarized in Fig. S2. The horizontal axis represents the prescribed maximum relative fabrication uncertainty. The vertical axis is $|\Delta S_{\max}/S_{\max}|$, where $\Delta S_{\max}$ is the maximum absolute difference between the corresponding elements of the perturbed $\mathbf{S}$ and the unperturbed $\mathbf{S}^{\mathrm{un}}$, while $S_{\max}$ is the maximum element magnitude of $\mathbf{S}^{\mathrm{un}}$. To provide a conservative estimate of fabrication tolerance, the maximum value of $|\Delta S_{\max}/S_{\max}|$ among the 1000 realizations is plotted for each perturbation level (Fig. S2).

As shown in Fig. S2, the normalized scattering-matrix deviation remains approximately proportional to the fabrication uncertainty over the entire perturbation range. Conventional CPA, 2nd-order and 3rd-order absorbing EPs exhibit very similar scaling behaviors. Increasing the EP order mainly leads to a moderate increase in the overall deviation level while preserving the same dependence on fabrication uncertainty. Therefore, although higher-order EPs are somewhat more sensitive to fabrication imperfections, no anomalous amplification of fabrication errors is observed. These results demonstrate that the proposed photonic-doped platform maintains a reasonable degree of robustness against fabrication variations.

## 6. Frequency-detuning analysis for the ZIM networks

To evaluate the influence of material dispersion and frequency detuning on the proposed ZIM network, we take the third-order absorbing EP configuration [Figs. 2(g)-2(i) of the Main Text] as a representative example for numerical studies.

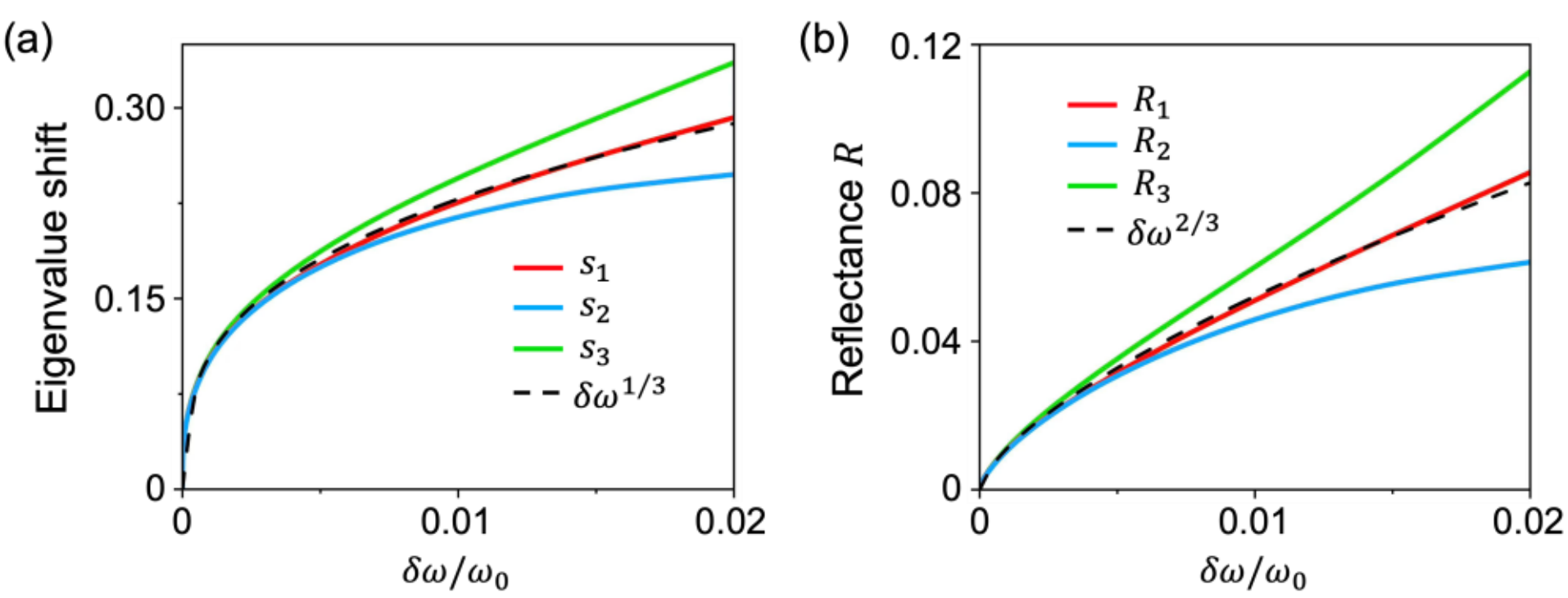

**Figure S3**. Frequency-detuning analysis for the third-order absorbing EP shown in Figs. 2(g)-2(i) of the Main Text. (a) Magnitudes of the three scattering-matrix eigenvalues as functions of the normalized frequency detuning $\delta\omega/\omega_0$. The black dashed line denotes the cube-root scaling. (b) Corresponding reflectance $R = |s|^2$, with the black dashed line denoting the two-thirds-power scaling.

To model realistic ZIM platforms, we assume linear dispersion: $\varepsilon = \mu = (\omega - \omega_0)/\omega_0$ for the central ZIM, $\mu = (\omega - \omega_0)/\omega_0$ for the two MNZ channels, and $\varepsilon = (\omega - \omega_0)/\omega_0$ for the ENZ channel. Here, $\omega_0$ denotes the central frequency at which the third-order absorbing EP is obtained. Such linear dispersion is typical for effective ZIMs based on photonic crystals [3, 4]. Full-wave simulations are then performed for frequency detuning $0 \leq \delta\omega/\omega_0 \leq$ $0.02$, and the corresponding scattering matrices are extracted to evaluate the eigenvalue evolution.

The results are presented in Fig. S3. Figure S3(a) shows the magnitudes of the three scattering matrix eigenvalues, while Fig. S3(b) shows the corresponding reflectance $R = |s|^2$. In the immediate vicinity of $\omega_0$, both eigenvalues and reflectance closely follow the characteristic third-order absorbing EP scaling $s \propto \delta\omega^{1/3}$ and $R \propto \delta\omega^{2/3}$, respectively. For larger frequency detuning, gradual deviations from the ideal $1/3$- and $2/3$-power laws are observed. Quantitatively, these fractional-power responses remain in good agreement with the ideal third-order EP behavior for frequency detuning up to approximately $|\delta\omega/\omega_0| \approx 0.3\%$. These results demonstrate that moderate frequency detuning does not immediately disrupt the EP-induced fractional-power response and confirm that the proposed platform retains a finite operational bandwidth.